\documentclass[twocolumn,showpacs,aps,amsmath,amssymb]{revtex4}

\usepackage{graphicx}

\def\a0{\alpha_0}
\def\a{\alpha}

\def\rtf{\rho_{\hbox{\tiny TF}}}
\def\rtfbar{\bar{\rho}_{\hbox{\tiny TF}}}

\def\r0{\rho_{0}}

\def\be{\begin{equation}}
\def\ee{\end{equation}}
\def\beq{\begin{equation}}
\def\eeq{\end{equation}}

\begin{document}

\title{Three-dimensional vortex structure of a
fast rotating Bose-Einstein condensate with harmonic-plus-quartic
confinement}

\author{Ionut Danaila}
\affiliation{Laboratoire
Jacques-Louis Lions,  Universit{\'e} Paris 6, 175 rue du Chevaleret,
75013 Paris, France.}
\date{\today}


\begin{abstract}
We address the challenging proposition of using real experimental
parameters in a three-dimensional (3D) numerical simulation of
fast rotating Bose-Einstein condensates. We simulate recent
experiments [V. Bretin, S. Stock, Y. Seurin and J. Dalibard, Phys.
Rev. Lett. 92, 050403 (2004); S. Stock, V. Bretin, S. Stock, F.
Chevy  and J. Dalibard, Europhys. Lett. 65, 594 (2004)] using an
anharmonic (quadratic-plus-quartic) confining potential to reach
rotation frequencies ($\Omega$) above the trap frequency
($\omega_\perp$). Our numerical results are obtained by
propagating the 3D Gross-Pitaevskii equation in imaginary time.
For $\Omega \leq\omega_\perp$, we obtain an equilibrium  vortex
lattice similar (as the size and number of vortices) to
experimental observations. For $\Omega>\omega_\perp$ we observe
the evolution of the vortex lattice into an array of vortices with
a central hole. Since this evolution was not visible in
experiments, we investigate the 3D structure of vortex
configurations and 3D effects on vortex contrast. Numerical data
are also compared to recent theory [D. E. Sheehy and L.
Radzihovsky, Phys. Rev. A  70, 063620 (2004)] describing vortex
lattice inhomogeneities and a remarkably good agreement is found.
 \end{abstract}

\maketitle

\section{Introduction}

In recent years, several experimental studies provided evidence
for the existence of quantized vortices in rotating Bose-Einstein
condensates (BEC's) \cite{mat,mad,mad2,abo,ram,hal,hod}. The
condensate is typically confined by a harmonic (quadratic)
potential with transverse frequency $\omega_\perp$ and starts to
nucleate vortices when the rotation frequency $\Omega$ exceeds a
critical value $\Omega_c$.
For increasing  $\Omega> \Omega_c$, more and more vortices appear
and arrange themselves into a regular triangular (Abrikosov)
lattice.

The fast rotation regime, corresponding to $\Omega \gtrsim
\omega_\perp$, is particularly interesting to explore since a rich
variety of scenarios are theoretically predicted: formation of
giant (multi-quantum) vortices, vortex lattice melting or quantum
Hall effects. This regime is experimentally delicate to
investigate \cite{ros} since for $\Omega=\omega_\perp$ the
centrifugal force compensates the trapping force and the
confinement vanishes. Using evaporative spin up, the Boulder group
has recently created condensates with rotation frequencies of the
order of $0.99 \omega_\perp$  and studied the properties of the
vortex lattice in the lowest Landau level \cite{eng,sch,cod}.

Another experimental approach to reach the fast rotation regime
was explored by the {\'E}cole Normale Sup{\'e}rieure (ENS) group
\cite{bre,sto,breth}. It consists in modifying the quadratic
trapping potential by superimposing  a blue detuned laser beam to
the magnetic trap holding the atoms. The resulting {\em
harmonic-plus-Gaussian} potential removes the singularity at the
limit $\Omega = \omega_\perp$ and allows to reach rotation rates
up to $\Omega \simeq 1.05 \omega_\perp$. The trapping potential
used in these experiments can be well approximated by a {\em
quadratic-plus-quartic} form, which has been extensively studied
lately \cite{fetter,lun,kas,kav,gho,jac,lundh,io2,kav2,fetstr}.
Different transitions involving a rich variety of vortex states
are theoretically predicted when $\Omega$ is increased: from a
dense vortex lattice to an array of singly quantized vortices with
a central hole and, finally, to a giant (multiquantum) vortex or
directly from a vortex lattice to a giant vortex.

For the highest rotation rates reached in experiments, neither
giant vortices nor vortex arrays with hole were clearly observed
\cite{bre,sto}. In exchange, a dramatic change in the appearance
of the condensate was reported: the vortices are less visible even
thought the gas remains ultracold and in fast rotation. The most
likely explanation for this intriguing behavior was related to the
transient character of the observed states leading to a fragile
vortex lattice dominated by three-dimensional (3D) effects
(vortices appear to have some excitation or bending leading to
poor optical contrast).

Since such effects are not trackable with previous (2D) numerical
approaches, the purpose of the present contribution is to
investigate the 3D structure of such condensates by numerically
generate the corresponding Gross-Pitaevskii (GP) wave function.
This is not without its challenges, since the description of a
prolate (cigar-shaped) condensate with a large number of vortices
(exceeding 100) requires very high spatial resolution and accurate
integration schemes. Computations become very expensive at high
rotation frequencies, which explains why such 3D simulations are
not, to the author's knowledge, currently available in the open
literature.

The numerically generated 3D condensates can be seen in
Fig.~\ref{fig-all}. For increasing rotation frequencies, the
vortex lattice evolves to a vortex array with a hole, which
confirms the scenario theoretically predicted
\cite{fetter,lun,gho,jac} and also observed in 2D simulations
\cite{kav,fetstr}. Since such transition was not observed in
experiments, we qualitatively analyze the obtained vortex states,
with a particular emphasize on the 3D features of vortex merging
leading to a central hole in the condensate.

Our analysis is then extended to quantitative comparisons to
experiments and theoretical predictions. We first check that
physical parameters (size, chemical potential) of numerical
condensates correspond well to available experimental ones. We
show in particular that the rotation frequencies reached in
experiments were not enough high to obtain an annular condensate.
We also measure from simulations the intervortex spacing and
compare the numerical results to recent theory of Sheehy and
Radzihovsky \cite{sheehy1,sheehy2} describing vortex lattice
inhomogeneities. A remarkably good agreement is found. Finally, we
discuss how 3D structure of vortices can affect optical contrast
of transient states observed in experiments.

\section{Physical parameters and numerical approach}

We consider a BEC of $N$ atoms confined by the trapping potential
$V$ and rotating along the $z$ axis at angular velocity
${\Omega}$. In the experiments at ENS \cite{bre,sto,breth},
$N=3{\times} 10^5$ atoms and the trapping potential can be written
as the superposition of the harmonic potential $V_h$ created by
the magnetic trap and the potential $U(r)$ introduced by the laser
beam propagating along the $z$ axis: \beq\label{eq-pot-trap}
V(r,z)= V_{h}(r,z)+U(r), \eeq with $r=\sqrt{x^2+y^2}$ and
\beq\label{eq-pot-exp} V_{h}={1\over 2}m
  (\omega^{(0)}_{\perp})^2 r^2+{1\over 2}m
  \omega^2_{z} z^2,
   \quad U(r)=U_0\, e^{ -2r^2/w^2}. \eeq
The trapping frequencies are $\omega^{(0)}_{\perp}=2\pi {\times}
75.5$ Hz and $\omega_{z}=2\pi {\times} 11$ Hz, resulting in a
cigar-shaped condensate. The laser waist is $w=25$ $\mu$m and the
amplitude of the laser beam is $U_0= k_B{\times} 90$ nK.

For $r/w$ sufficiently small, the potential $V(r)$ can be
approximated by: \beq\label{eq-pot-approx} V_{1}= \left[{1\over
2}m
  (\omega^{(0)}_{\perp})^2 -{{2U_0}\over w^2}\right] r^2 +{{2U_0}\over
  w^4}r^4 + {1\over 2}m
  \omega^2_{z} z^2.\eeq
For this {\em quadratic-plus-quartic} potential, the transverse
trapping frequency is decreased to $\omega_\perp=2\pi {\times}
65.6$ Hz. Since the amplitude $U_0$ of the laser beam is low in
experiments, the quadratic part of the potential $V_1$ remains
positive (repulsive interactions) and the quartic part is very
small. It is interesting to note that a stronger amplitude $U_0$
could generate a {\em quartic-minus-quadratic} potential, which
was theoretically studied in Refs. \cite{io2,jac,gho2}.

\begin{figure*}[t]
\centerline{\includegraphics[width=0.95\textwidth]{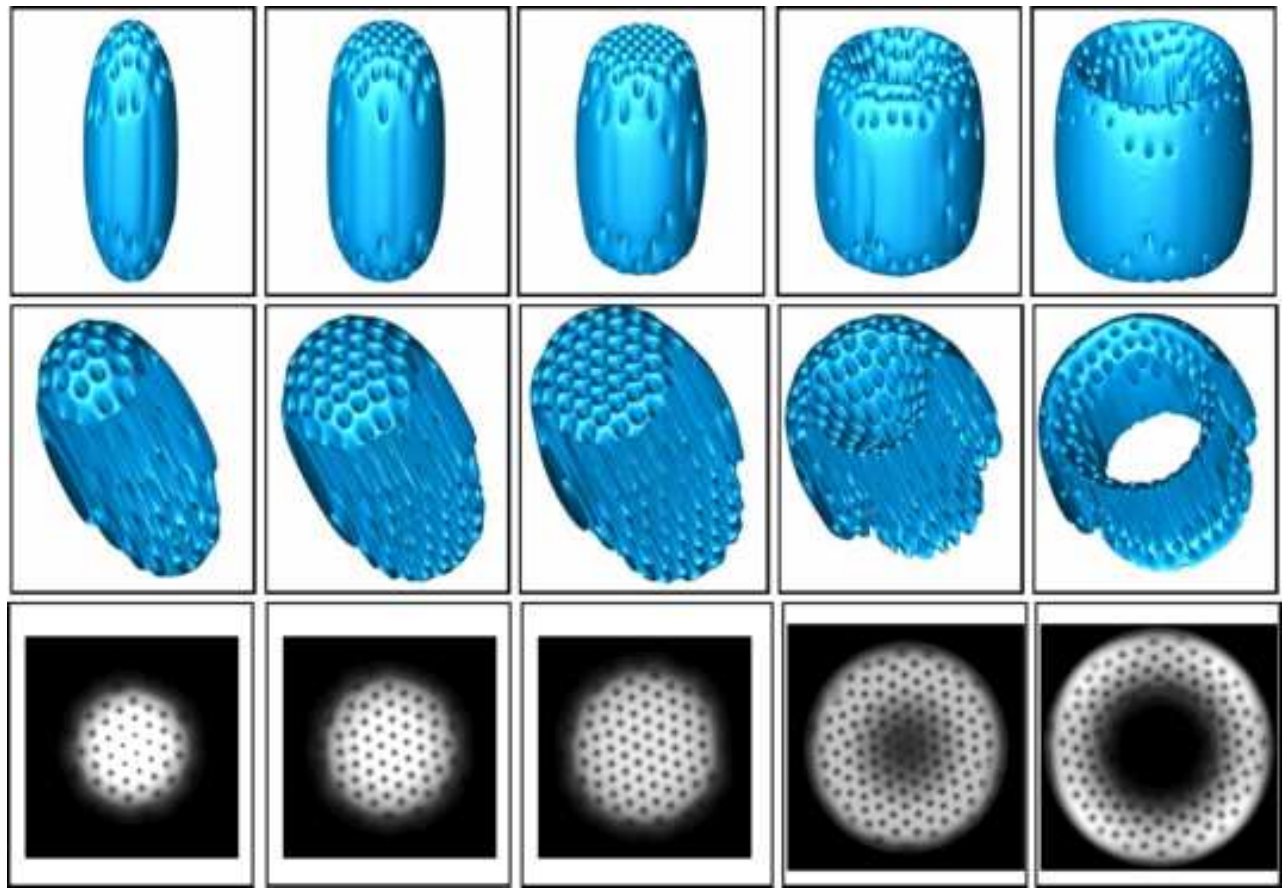}}
 \caption{(Color online) Numerically generated condensates obtained using
a {\em quadratic+quartic} trapping potential with the parameters
corresponding to experiments of \cite{bre,sto}. Each column
corresponds to a value of the rotation frequency - from left to
right: $\Omega/2\pi=60, 64, 66, 70.6, 73$ (respectively,
$\Omega/\omega_\perp=0.92, 0.98, 1.01, 1.08, 1.11$).
Three-dimensional views of the vortex lattice identified by means
of iso-surfaces of low atomic-density (first two rows) and
contours of density integrated along the rotation ($z$) axis. Note
that the formation of the hole in the condensate is not complete
for $\Omega/2\pi=70.6$ and we still distinguish individual
singly-quantized vortices in the center (see also Fig.
\ref{fig-bones} for a detailed picture of this configuration).}
 \label{fig-all}
\end{figure*}

The numerical results presented in this paper were obtained using
a {\em quadratic-plus-quartic} potential (Eq.
\ref{eq-pot-approx}), for which extensive theoretical studies are
available \cite{fetter,lun,kas,kav,gho,jac,lundh,io2,kav2,fetstr}.
Numerical simulations using the {\em quadratic-plus-Gaussian}
original potential (Eq. \ref{eq-pot-exp}) showed the same
qualitative evolution of the vortex configuration as in Fig.
\ref{fig-all}, with a transition to a vortex array with hole for a
slightly lower rotation frequency.

As a numerical approach, we compute the macroscopic wave-function
$\psi(x,y,z)$  by propagating the three-dimensional
Gross-Pitaevskii (GP) equation in imaginary time by the numerical
method used in Refs. \cite{io1,io2,io3}. After rescaling the GP
equation as in Ref. \cite{AR}, a hybrid Runge-Kutta-Crank-Nicolson
scheme is used for the time integration and a sixth-order compact
finite difference scheme for the space discretization.

As initial condition we generally use a vortex-free density
distribution following the Thomas-Fermi (TF) law:
 \begin{equation}
\rtf(r,z)=\frac{m}{4\pi \hbar^2 a_s}\left(\mu- V_{1}(r,z)+{1\over
2}m\Omega^2 r^2 \right)\ ,
 \label{eq-tf}
 \end{equation}
where $a_s=5.2$~nm is the scattering length and $\mu$ the chemical
potential given by the constraint $\int d^3r \rtf =N$. For the
{\em quadratic-plus-quartic} trapping potential $V_1$, an exact
analytical form for $\mu$ can be derived \cite{fetter} depending
on the value of $\Omega$ which dictates the shape of the
condensate (with or without a hole). The maximum transverse radius
$R_\perp$ and longitudinal half-length $R_z$ can be then
calculated from Eq. (\ref{eq-tf}) in order to estimate the
dimensions of the rectangular computational domain. For high
$\Omega$ (when the condenstae is nearly spherical and more than
100 vortices are present), up to $240\times 240 \times 240$ grid
points are used to compute equilibrium states.

%

The post-processing of the results follows the experimental
approach \cite{bre,cod} (with the difference that the radial
expansion after the time of flight is not numerically simulated).
The numerical 3D wave-function is converted to an atomic density
$\rho(x,y,z)=|\psi(x,y,z)|^2$ and integrated along the rotation
($z$) axis. The resulting 2D-density $\bar{\rho}^z(x,y)$
(isocontours are displayed in Fig. \ref{fig-all}, last row of
images) will be used in the following for comparison to
experiments and theory.

\section{Description of the results}

The evolution of the 3D structure of the condensate with
increasing $\Omega$ can be seen in Fig. \ref{fig-all}. We start
with a qualitative description of vortex configurations.  The
obtained results will be then analyzed quantitatively and compared
to available experimental and theoretical values. All quantitative
parameters discussed in this paper are summarized in table
\ref{tab-char}.

\begin{table}[!h]
  \begin{tabular}{|c|c|c|c|c|c|}\hline
$\Omega/(2\pi)$ & 60 & 64 & 66 & 70.6 & 73\\
$\Omega/\omega_\perp$ &0.92 &0.98& 1.01& 1.08 &1.11\\ \hline
\hline
 $R_\perp [{\mu}m]$ & 10.4&12.2 & 13.2 & 17.2 & 19.2
\\ \hline
 $R_z [{\mu}m]$ & 29.0&25.4 & 22.5 &  20.1 & 18.6
\\ \hline\hline
 $N_v$
&37& 51& 62& 126 & 113 \\ \hline
 $L_z
[\mbox{units of $\hbar$}]$ &17.4&28.5&39.1&122.6&239.1 \\
\hline\hline
 $r_v/\xi$ & 2.15 & 1.84 & 1.65 & 1.36 & 1.76\\ \hline
\end{tabular}
\caption{Summary of the characteristics of numerically generated
condensates: (maximum) transverse radius $R_\perp$ and
longitudinal half-length $R_z$; number of vortices $N_v$ and
angular momentum $L_z=i \int\,d^3r \bar{\psi} \left( y{\partial
\psi}/{\partial x} - x {\partial \psi}/{\partial y}\right)$;
scaling constant for the ratio between vortex-core radius $r_v$
and healing length $\xi$ [obtained from integrated density
$\bar{\rho}^z(x,y)$].} \label{tab-char}
\end{table}

\subsection{Vortex configurations}

For rotation frequencies below $\omega_\perp$ ($\Omega/(2\pi)=60$
and $64$) the condensate has the usual prolate shape (see Fig.
\ref{fig-all}, first two columns). Vortices near the center of the
condensate are straight and form a regular triangular lattice.
Vortices located near $r=R_\perp$ are bending, reaching the
surface of the condensate using the shortest path. These outer
vortices are not symmetrically arranged and have different
lengths. It is interesting to note that for these two values of
$\Omega$, the number of vortices $N_v$ we find numerically
($N_v=37$ and $51$) is very close to that visible in experimental
pictures \cite{bre} ($N_v^{expt}=30$ and $52$).

Starting with  $\Omega/(2\pi)=66$ ($\Omega/\omega_\perp=1.01$),
the experimental pictures show a lack of contrast for entire zones
of the vortex lattice. Vortices are less visible and do not allow
a proper estimation of the rotation frequency from vortex surface
density. Numerical condensate for this rotation frequency (Fig.
\ref{fig-all}, third column) display a well-defined triangular
vortex lattice. Most of the vortices are straight and join the top
and bottom ends of the condensate which are almost flat. This
particular shape of the condensate corresponds well to that
predicted from the TF law (\ref{eq-tf}). Indeed, for
$\Omega=\omega_\perp$, the density distribution $\rtf(r,z)$
depends only on the quartic part of the trapping potential $V_1$
and the surface  of the condensate defined as $\{\rtf= 0\}$  is
flat near the rotation $z$ axis.

For rotation frequencies exceeding $\omega_\perp$, experimental
condensate exhibits a local minimum in the central density, but
the theoretically predicted \cite{kav,fetter} transition to a
vortex lattice with a hole (annular condensate) is not
experimentally reported. This is the case in our simulations (Fig.
\ref{fig-all}). The rotation frequency corresponding to this
transition is found to be $\Omega_h/(2\pi)=71$, a value close to
the TF prediction $\Omega_h^{\hbox{\tiny TF}}/(2\pi)=70$. These
values are already larger than those attained in experiments
($\Omega/(2\pi)< 69$), which can simply explain why the hole was
not experimentally observed.

\begin{figure}[!h]
\centerline{\includegraphics[width=0.7\columnwidth]{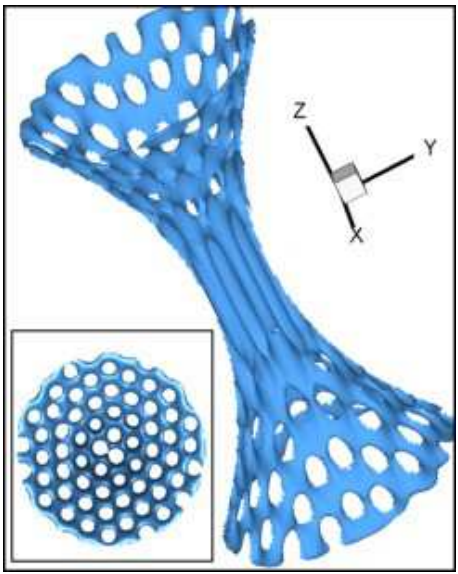}}
 \caption{(Color online) Details of the vortex configuration for $\Omega/(2\pi)=70.6$.
 Vortices near the rotation axis are isolated, showing the merging
 process that will finally lead to
the formation of a central hole. Insert: top view of the same
configuration.} \label{fig-bones}
\end{figure}

The numerically generated condensates before and after transition
to an annular condensate are shown in Fig. \ref{fig-all} (last two
columns of images). For $\Omega/(2\pi)=70.6$, the central hole is
not yet formed since the top and bottom depletions have not
merged. At the center of the condensate, the density is very low
but not zero, and we can still distinguish individual vortices
from isocontours of the density integrated along the $z$ axis
(Fig. \ref{fig-all}). Since the contrast in this last image is low
near the center, we show details of the vortices near the rotation
axis in Fig. \ref{fig-bones}. In the center there are three
vortices with larger cores that start to reconnect at the top and
bottom of the condensate. This merging process is highly
three-dimensional and will finally lead to the formation of a
central hole for higher $\Omega$.

The structure of the condensate is completely different for
$\Omega/(2\pi)=73$ (last column of images in Fig. \ref{fig-all}).
The condensate is nearly spherical, with a large central hole
surrounded by three concentric circles of singly quantized
vortices. Most of the 113 identified vortices are bent, reaching
either inner or outer faces of the condensate. Since convergence
for this case is particularly slow (two weeks of computational
time is necessary using a PC workstation), we did not explore
cases for higher $\Omega$. For the considered parameters, a second
transition to a configuration with a pure giant vortex (without
singly quantized vortices in the annular region) may occur at very
high rotation frequencies \cite{fetstr} that are not numerically
affordable in 3D.

\subsection{Vortex lattice inhomogeneity}

We now turn on more quantitative analysis of numerical results.
Before analyzing the characteristics of the vortex lattice, we
first check that the dimensions of the numerical generated
condensates correspond well to experimental ones. The density
$\bar{\rho}^z$ is integrated along the azimuthal direction
$\theta$ to get the radial density profile
$\bar{\rho}^{z,\theta}(r)$. This  profile is fitted to the
Thomas-Fermi distribution (\ref{eq-tf}), taking the chemical
potential $\mu$ and the rotation frequency $\Omega$ as adjustable
parameters. The theory fit value of $\Omega$ is within 1\% of the
value of $\Omega$ for which the computation was done.

\begin{figure}[!h]
\centerline{\includegraphics[width=\columnwidth]{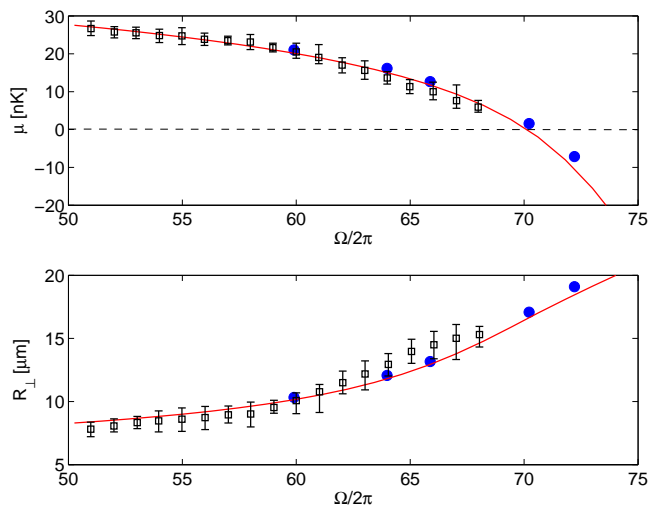}}
 \caption{(Color online) Chemical potential ($\mu$) and maximum transverse
  radius of the condensate ($R_\perp$)
  as functions of the rotation frequency. Experimental measurements from
   Ref. \cite{breth} (squares),
  numerical results (circles) and Thomas-Fermi theoretical
  prediction (solid line).} \label{fig-cpot}
\end{figure}

The resulting chemical potential $\mu$ and the transverse radius
$R_\perp$ (which is the maximum radius for the condensate with
hole) are compared in Fig. \ref{fig-cpot} to experimental values
from Ref. \cite{breth} and Thomas-Fermi approximation
(\ref{eq-tf}). For the experimentally available range of rotation
frequencies, numerical results are in good agreement with
experimental and theoretical values. For values of $\Omega$ not
available experimentally, numerical results follow the TF
prediction. In particular, the numerical value
$\Omega_h/(2\pi)=71$ for which the central hole first appears in
the condensate (corresponding to a chemical potential $\mu=0$) is
well predicted by the TF law ($\Omega_h^{\hbox{\tiny
TF}}/(2\pi)=70$).

We continue our dimensional analysis by extracting the
characteristics of the vortex lattice: namely the intervortex
spacing $b_v$ and the vortex core size $r_v$. We follow a similar
post-processing procedure as in Ref. \cite{cod}. Using the
integrated (along $z$) density field $\bar{\rho}^z(r,\theta)$, we
identify vortex centers by 2D searching of local minima. Resulting
points are checked to correspond to vortices visible in Fig.
\ref{fig-all} (last row of images). Assuming a triangular lattice
structure, we select vortices for which the six nearest vortex
neighbors form a hexagonal pattern. Only for such vortices (i.e.
vortices close to $R_\perp$ are discarded), is the intervortex
spacing $b_v$  measured by averaging the distance from the vortex
center to the centers of the six neighbors. The vortex core radius
$r_v$ is measured as follows: for a given vortex located at
$(r_0,\theta_0)$, the density profile $\bar{\rho}^z_v(r)$ along
the radius passing through the center of the vortex is extracted
from the 2D field $\bar{\rho}^z$; by subtracting the integrated TF
density profile ${\rtfbar}^z(r)$ [corresponding to Eq. \ref{eq-tf}
integrated along $z$], we obtain a vortex-core residual that is
fitted with a Gaussian profile:
\begin{equation}
{\rtfbar}^z(r)-\bar{\rho}^z_v(r)=A \exp\left[-\frac{1}{2}(r-{
r_0})^2/{r_v}^2\right].
\end{equation}
The amplitude $A$ is used to define the vortex contrast \cite{cod}
as $A/{\rtfbar}^z(r_0)$, i.e. the ratio between the "missing"
column density at vortex center $r_0$ and the corresponding TF
value. Only vortices with a contrast greater than 0.7 are
considered to compute core radii $r_v$.

\begin{figure}[h!]
\centerline{\includegraphics[width=0.96\columnwidth]{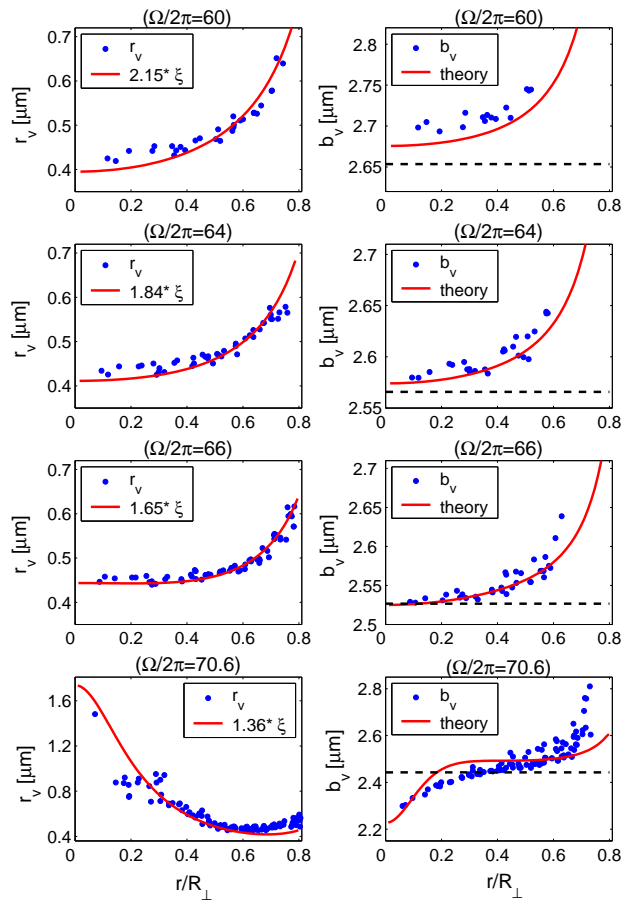}}
 \caption{(Color online) Variation  of vortex core radius $r_v$ and
 intervortex spacing $b_v$  (values in $\mu$m) as functions of the
  non-dimensional radius $r/R_\perp$. For each plot, the value of the rotation frequency
($\Omega/2\pi$) is indicated. In plots displaying $r_v$, solid
line represents the variation of the healing length $\xi$, scaled
by a constant indicated in the legend. Variation of $b_v$ is
compared to theory prediction of Sheehy and Radzihovsky
\cite{sheehy1,sheehy2} (solid line) and the estimation assuming a
uniform (rigid-body) vortex distribution (dashed line).}
 \label{fig-rvbv}
\end{figure}

Figure \ref{fig-rvbv} shows the variation  of $r_v$ and $b_v$ as
functions of the non-dimensional radius $r/R_\perp$. Values are
given in $\mu$m and rotation frequencies $\Omega/\omega_\perp \leq
1.01$ are considered (condensates without hole). As expected
\cite{fetter,cod}, the core radius $r_v$ scales with healing
length, defined from the TF density fit
 $\xi(r)=[8\pi a_s {\rtfbar}^z(r)]^{-1/2}$. The scaling constant
(also summarized in Tab. \ref{tab-char}) decreases with $\Omega$,
with values comparable to those found in Ref. \cite{cod} for a
harmonic trapping potential. We recall that the values presented
here correspond to a post-processing for $r_v$ using integrated
density ${\bar{\rho}}^z$, as in experiments. A similar
post-processing using the 2D density field $\rho$ extracted from
the 3D simulation at $z=0$, revealed scaling constants for
$r_v/\xi$ of order of 1 (more precisely, $r_v/\xi \approx 0.98,
0.93, 0.86$ for, respectively, $\Omega/2\pi=60, 64, 66$).

The calculated intervortex spacing $b_v$ is compared in Fig.
\ref{fig-rvbv} to recent theory of Sheehy and Radzihovsky
\cite{sheehy1,sheehy2}. They expressed the vortex density $n_v(r)$
as a function of the local superfluid density $\rho_s(r)$:
\begin{equation}
n_{v}(r) = \frac{\Omega m}{\pi\hbar}+ ln[\hbar/(2.718\,m
\Omega\xi_v^{2})]\, \nabla^2\left[ln(\rho_s(r))\right].
\label{eq-nv}
\end{equation}
The second term in (\ref{eq-nv}) is a small correction to the
vortex density for a uniform vortex distribution corresponding to
a rigid-body rotation $n_{v0}=({\Omega m})/({\pi\hbar})$. The
vortex density $n_v$ can be converted to intervortex spacing by:
\begin{equation}
    b_v(r) = \sqrt{2/(3^{1/2} n_{v}(r))}.
\end{equation}

Numerical results are compared to theoretical predictions using in
Eq. \ref{eq-nv}  the TF fit for the integrated density profile
($\rho_s(r)={\rtfbar}^z(r)$) and the characteristic length for the
vortex-core $\xi_v$ defined as \cite{sheehy1}
$\xi_v=\hbar/(m\omega_\perp R_\perp)$. The agreement is remarkably
good. For $\Omega/2\pi\leq 64$, the density profile is close to an
inverted parabola (the influence of the quartic term being small)
and $b_v$ is monotonically increasing with $r$. Similar results
were reported for a harmonic trapping potential \cite{cod}. As
expected, the estimation using the rigid-body rotation assumption
(dashed line in the plot) becomes better with increasing $\Omega$
(the lattice becomes denser). For $\Omega/2\pi=70.6$, the density
profile has a Mexican-hat structure and vortices are constrained
to agglomerate towards the center, where density is small. The
intervortex spacing is small near the center and increases to the
rigid-body value near $r/R_\perp\simeq 0.5$ where the density is
maximum. The theory nicely illustrates this complex dependance of
$b_v$ on the radial position.

\section{Discussion and conclusion}

We have presented in this paper three-dimensional numerical
results for a fast-rotating BEC trapped in quadratic-plus-quartic
potential corresponding to experiments at ENS \cite{bre,sto}. The
obtained vortex configurations show a transition from a dense
vortex lattice to a vortex array with a central hole at a critical
rotation frequency $\Omega_h/(2\pi)=71$. This result confirms
theoretical and 2D numerical results \cite{fetter,kav,fetstr} and
goes beyond experimental observations, since experiments failed to
reach rotation frequencies close to $\Omega_h$.

Our results also support the assumption \cite{bre} that vortices
are less visible in experiments for $\Omega/(2\pi)\ge 66$ because
of the fragility of the vortex lattice which becomes dominated by
3D-effects, such as vortex bending. In order to illustrate this
statement it is worth describing how the condensate evolves in the
"imaginary" time (i.e. how it relaxes to an equilibrium state).

The imaginary-time evolution of the condensate looks similar to a
real-time evolution. When suddenly increasing $\Omega$, new
vortices are generated at the border of the condensate and enter
the condensate. In the first stages of the computation, 3D vortex
lines are strongly distorted, giving a {\em spaghetti} image of
the lattice. Close to equilibrium, vortices become straight in
their central part and arrange themselves in a more and more
regular lattice. Convergence is particularly slow at the end of
the computation when the position and  shape of vortices evolve
very slowly. Convergence is considered when the energy remains
constant (relative fluctuations less than $10^{-6}$) for a
relatively long time to be sure that a stable state was obtained.
The convergence time is much longer (roughly by a factor of 2) for
values of the rotation frequency exceeding $\omega_\perp$.

\begin{figure}[h!]
\centerline{\includegraphics[width=\columnwidth]{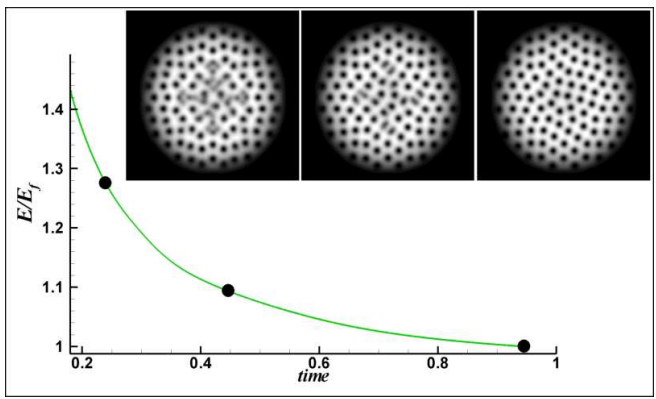}}
 \caption{Example of  energy decrease during the propagation of
3D Gross-Pitaevskii equation in imaginary time. Simulation for
$\Omega/(2\pi)=66$, using the quadratic-plus-Gaussian trapping
potential (\ref{eq-pot-exp}). Energy is normalized by the
equilibrium (final) value $E_f$. Inserts show iso-contours of the
integrated (along $z$) density corresponding to three successive
time instants
represented on the energy curve.}
 \label{fig-contrast}
\end{figure}

An example of intermediate states of the condensate before
reaching a converged equilibrium state is displayed in Fig.
\ref{fig-contrast}. The simulation corresponds to a
quadratic-plus-Gaussian trapping potential (\ref{eq-pot-exp}),
(closer to the experimental one) and a vortex configuration
without hole. Transient states look closer to experimental
pictures than the equilibrium states presented in Fig.
\ref{fig-all}. Three-dimensional exploration of the condensate
reveals that vortices which are less visible have distorted
structures which diminish the contrast in an integrated view along
the $z$ axis. These effects are stronger for condensates
displaying a central depletion; even for equilibrium states of
such condensates, it is difficult to distinguish individual
vortices in the center, as can be seen in Fig. \ref{fig-all} for
$\Omega/(2\pi)=70.6$. This confirms the hypothesis \cite{bre} of
the fragility of the experimental vortex lattice for high rotation
frequencies: for transient states, 3D vortex lines have some
excitations leading to a poor optical contrast. It is possible
that the very low temperature in experiments slows down the
dissipative process allowing only the observation of transient
states dominated by 3D effects. But is not to be excluded that
 a thermal excitation may be at the origin of
the vortex-line bending responsible for low optical contrast and, therefore,
increasing the temperature in experiments is not a solution to improve vortex
lattice contrast.

Our simulations also offer a detailed 3D picture of vortex
configurations that is not available from experiments and 2D
simulations. In particular, the vortex merging leading to the
formation of the central hole in a condensate is proved to be
highly three-dimensional. Quantitative measurements of the
intervortex spacing give a new validation of the theoretical study
of Sheehy and Radzihovsky \cite{sheehy1,sheehy2} predicting vortex
lattice inhomogeneity from local density profile. An interesting
question remaining for future numerical investigations is whether
or not the condensate trapped in a quadratic-plus-quartic
potential enters the lowest Landau level regime for
$\Omega\simeq\Omega_h$.

\noindent {\bf Acknowledgments:} I am grateful to J. Dalibard and
S. Stock for helpful comments on the manuscript and for allowing
me to use their experimental data. A. Aftalion and V. Bretin are
also acknowledged for useful discussions.

\end{document}